\documentclass{article}

\def\xxinput#1{\input#1}

\xxinput{vsolj02.sty}

\usepackage[dvipdfmx]{graphicx}
\usepackage[comma,colon]{natbib}

\usepackage[OT2,T1]{fontenc}

\def\cite{\citealt}
\setcitestyle{aysep={}}

\newcounter{author}
\def\altaffilmark#1{$^{#1}$}
\def\altaffiltext#1{$^{#1}$\,}

\def\authorcount#1#2{{\refstepcounter{author}\label{#1}
                     \altaffiltext{\ref{#1}}{#2}}}

\begin{document}

\begin{center}

\title{Gaia22ayj: outburst from a deeply eclipsing 9.36-min binary?}

\author{
        Taichi~Kato\altaffilmark{\ref{affil:Kyoto}}
}
\email{tkato@kusastro.kyoto-u.ac.jp}

\authorcount{affil:Kyoto}{
     Department of Astronomy, Kyoto University, Sakyo-ku,
     Kyoto 606-8502, Japan}

\end{center}

\begin{abstract}
\xxinput{abst.inc}
\end{abstract}

  Gaia22ayj is a transient detected by the Gaia Photometric
Science Alerts Team\footnote{
  $<$http://gsaweb.ast.cam.ac.uk/alerts/alert/Gaia22ayj/$>$.
}.
The object was detected at 17.01 mag on 2022 March 3.
Five Gaia observations between March 3.112 and 4.038 UT
showed significant short-term variations between
16.55 and 17.01.  The object had been registered as
a variable ZTF19aagmvuk\footnote{
  $<$https://lasair.roe.ac.uk/object/ZTF19aagmvuk/$>$.
} by the Zwicky Transient Facility (ZTF: \cite{ZTF}).
The same outburst was detected in outburst by ZTF and
the Transient Alert Broker Lasair \citep{ZTFLasair}
reported $r$=16.09 on March 3.399 UT.
The Gaia EDR3 position, magnitudes and parallax
08$^{\rm h}$ 25$^{\rm m}$ 26\hbox{$.\mkern-4mu^{\rm s}$}528,
$-$22$^{\circ}$ 32$^\prime$ 12\hbox{$.\mkern-4mu^{\prime\prime}$}34
(J2000.0), $BP$=19.570, $RP$=18.461 and $\varpi$=0.43(22) mas,
respectively \citep{GaiaEDR3}.

\begin{figure*}
\begin{center}
\includegraphics[width=16cm]{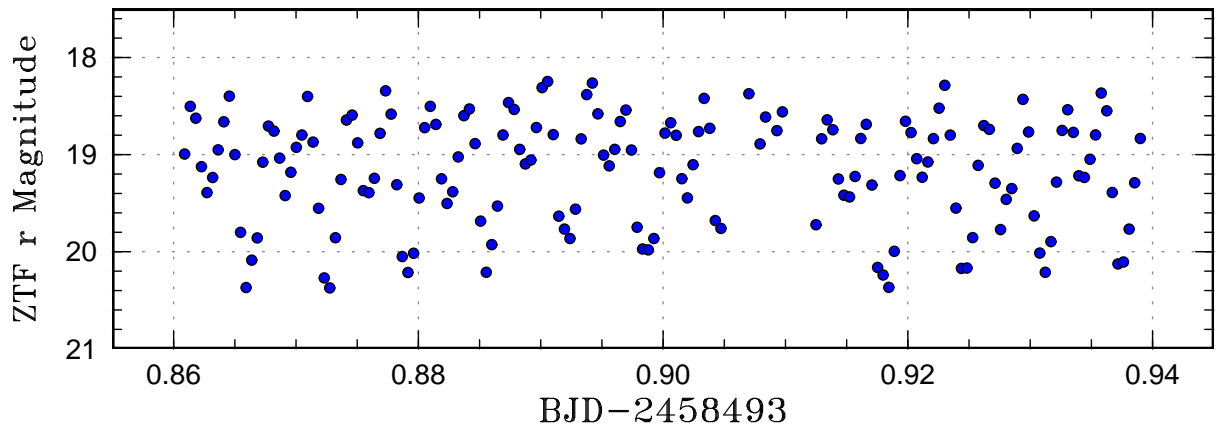}
\caption{
  ZTF light curve of Gaia22ayj on 2019 Jan. 10.
  Periodic short-period variations are clearly seen.
}
\label{fig:lc}
\end{center}
\end{figure*}

\begin{figure*}
\begin{center}
\includegraphics[width=16cm]{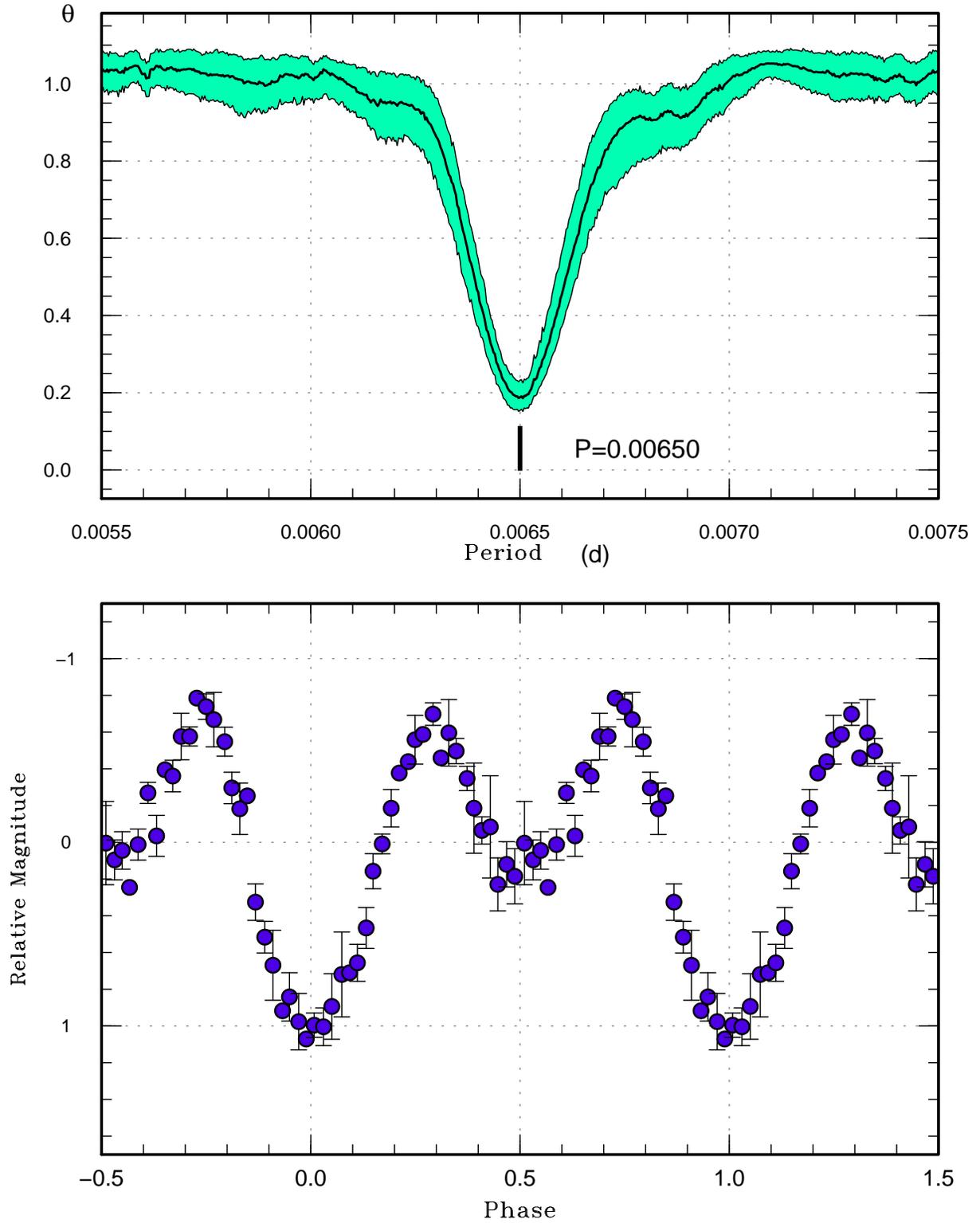}
\caption{PDM analysis of Gaia22ayj using the ZTF data
  on 2019 Jan. 10.
  (Upper): PDM analysis.
  (Lower): mean profile.
  The epoch and period were from equation (\ref{equ:eph}).
}
\label{fig:pdm}
\end{center}
\end{figure*}

\begin{figure*}
\begin{center}
\includegraphics[width=16cm]{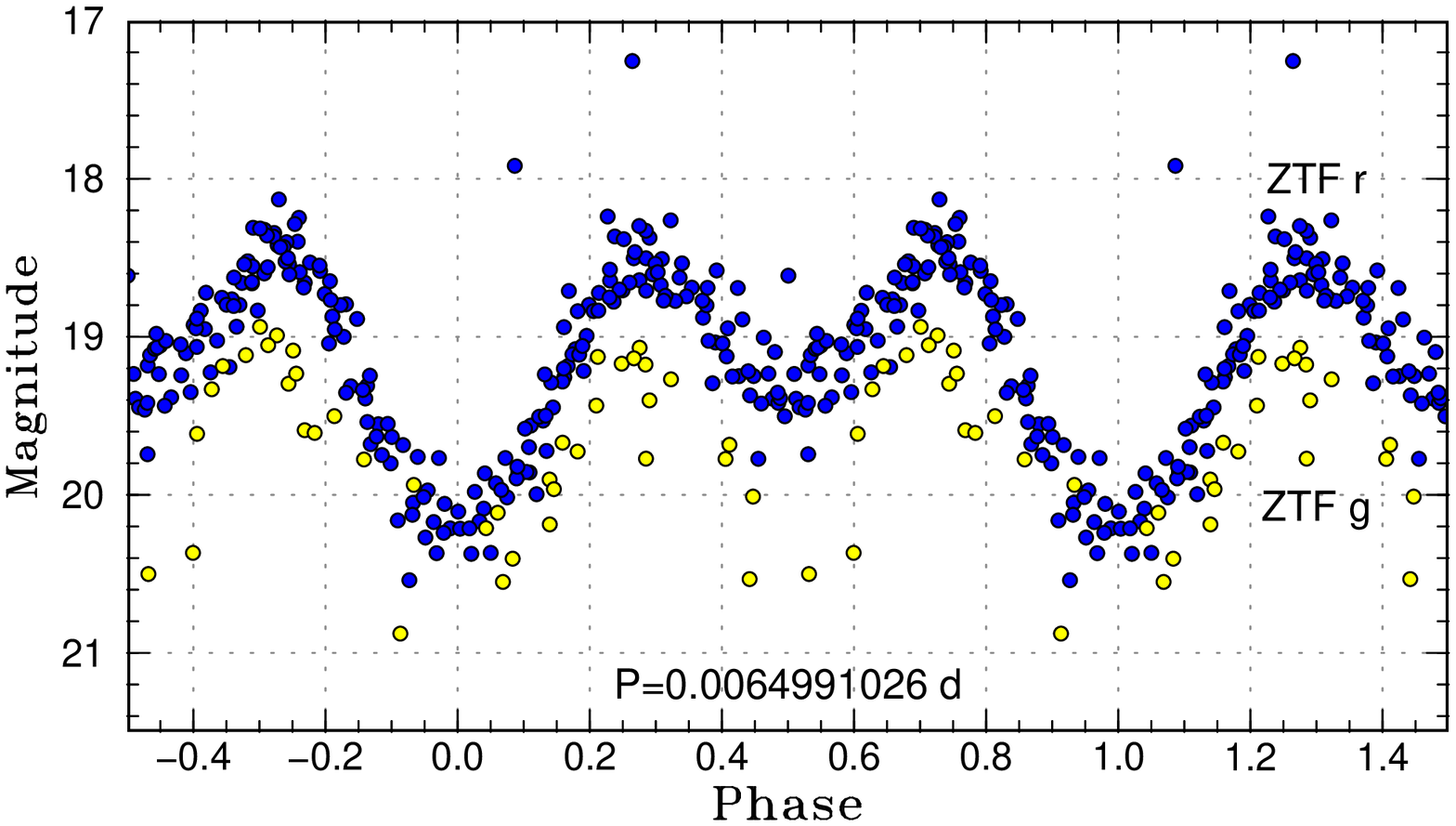}
\caption{The phase-folded light curve of the entire ZTF data.
  The epoch and period were from equation (\ref{equ:eph}).
}
\label{fig:phase}
\end{center}
\end{figure*}

  Using the ZTF public data\footnote{
  The ZTF data can be obtained from IRSA
$<$https://irsa.ipac.caltech.edu/Missions/ztf.html$>$
using the interface
$<$https://irsa.ipac.caltech.edu/docs/program\_interface/ztf\_api.html$>$
or using a wrapper of the above IRSA API
$<$https://github.com/MickaelRigault/ztfquery$>$.
}, I found coherent short-period variations
in time-series ZTF observations on 2019 Jan. 10
(T. Kato, vsnet-alert 26674\footnote{
   $<$http://ooruri.kusastro.kyoto-u.ac.jp/mailarchive/vsnet-alert/26674$>$.
}; figure \ref{fig:lc}).
Phase dispersion minimization (PDM: \cite{PDM}) analysis
yielded a period of 0.006500(7)~d with two different
maxima and minima in one cycle (figure \ref{fig:pdm}).
I have confirmed that this period can express
the entire ZTF light curve as well.
The orbital period was refined using
the Markov-Chain Monte Carlo (MCMC)-based
method introduced in \citet{Pdot2}.
The resultant ephemeris is
\begin{equation}
{\rm Min\;I\;(BJD)} = 2458699.17927(1) + 0.00649910257(13) E.
\label{equ:eph}
\end{equation}
The phase-folded light curve of the entire ZTF data
(excluding the Lasair data) is shown in
figure \ref{fig:phase}.  This period express all the
observations very well.

\begin{figure*}
\begin{center}
\includegraphics[width=16cm]{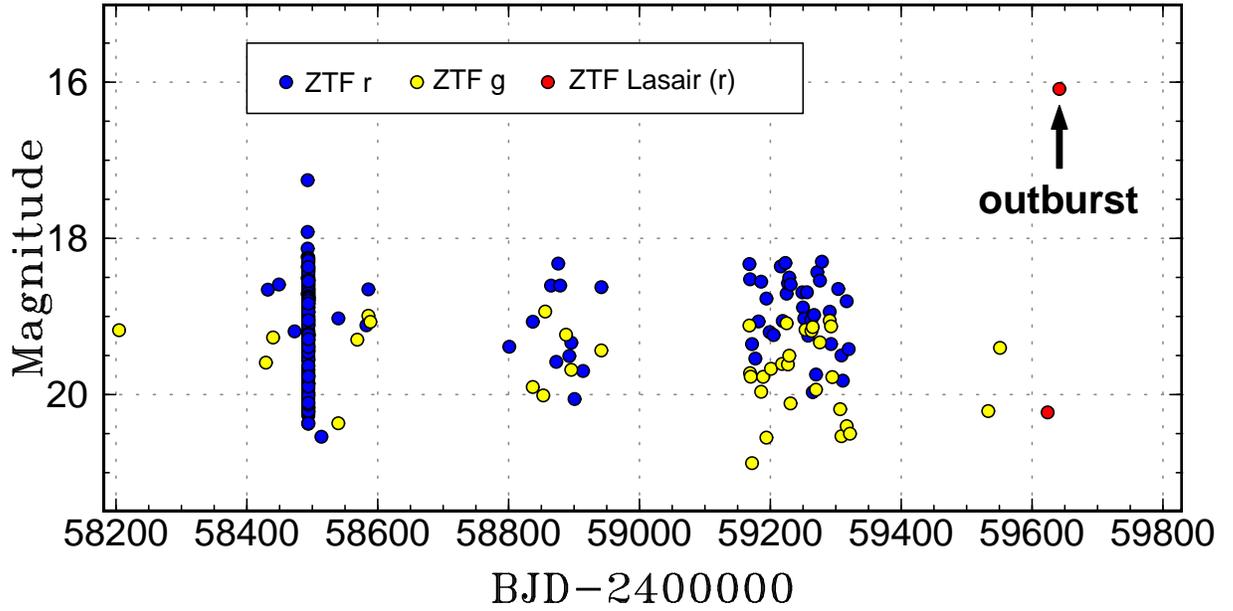}
\caption{
  Long-term ZTF light curve of Gaia22ayj.
  An outburst was detected on 2022 March 3.
}
\label{fig:long}
\end{center}
\end{figure*}

  The resultant period of 0.0064991026~d = 9.36~min
is very short.  Although this could be the spin period
of a binary containing a white dwarf, such as in
the white dwarf pulsar AR Sco
\citep{mar16arsco,sti18arsco} and
candidate white dwarf pulsars
ZTF J185139.81$+$171430.3 = ZTF18abnbzvx
\citep{kat21j185139,kli21j1851atel14965,pav21j1851atel14973},
ASASSN-V J205543.90$+$240033.5
\citep{kat21j205543,kat21j205543rev,wag21j2055},
and more recently GLEAM-X J162759.5$-$523504.3
\citep{kat22j1627}.
The profile of variations in Gaia22ayj
and the constancy of the profile, however, do not look
similar to those of white dwarf pulsars.

  The phase-folded light curve rather suggests
a $\beta$ Lyr-type or W UMa-type eclipsing binary with
different minima.  The slightly brighter second maximum
is suggestive of the O'Connell effect \citep{oco51effect}.
The only such short-period
(9.36~min) binaries known to us are AM CVn stars
composed of white dwarfs [For a review of AM CVn stars
see e.g. \citet{sol10amcvnreview}].
Among AM CVn stars, two short-period systems
HM Cnc (5.36~min) and V407 Vul (9.48~min) are
considered to be direct impact accretors without
forming an accretion disk
\citep{mar02v407vul,ram02j0806}.
The third shortest system ES Cet (10.3~min)
has recently been shown to have an accretion disk
\citep{bak21escet}.
Aside from these accreting systems, an eclipsing binary
with a period of 6.91~min (ZTF J153932.16$+$502738.8)
is also known \citep{bur19j1539}.

  If Gaia22ayj is indeed an accreting 9.36-min eclipsing
binary, this object is the shortest period known such
an object (the record up to now was ES Cet).  The presence
of eclipses in ES Cet was confirmed only by
rotational disturbance in the profiles of emission lines
\citep{bak21escet}, and not yet by photometry.
Gaia22ayj would be very important case due to
the very large amplitudes of photometric variations.
Following the interpretation of an eclipsing binary,
the large depths of the eclipses require that the system
is seen almost edge-on and the brightness of the two
components do not differ greatly.  A simplified
calculation suggests that a luminosity ratio of $\sim$0.4
could reproduce the observed difference in depths of
the minima.  With this orbital period, the Doppler
beaming is expected to be an order of 1\%(=0.01~mag)
\citep{zuc07beaming} and would be detectable by
high-precision photometry.

  The presence of an outburst lasting nearly 1~d in
Gaia22ayj suggests the presence of instability in
the accretion disk, rather than a short-lived flare or
an accretion event as seen in in intermediate polars.
The detection in X-rays by the Swift X-ray telescope
(1SWXRT J082526.4$-$223212: \cite{del13SWIFTcat})
is compatible with an accreting binary.
The presence of an accretion disk makes Gaia22ayj
similar to ES Cet.  The disks in short-period AM CVn
stars, like ES Cet, are considered to be thermally stable
due to the high mass-accretion rate
considering the standard evolutionary path
in which the mass transfer is driven by gravitational
wave radiation and the mass-losing white dwarf
fills the Roche lobe
\citep{tsu97amcvn,kot12amcvnoutburst}.
The case of Gaia22ayj appears to be different.
If Gaia22ayj is indeed an AM CVn-like object,
it might be in the turn-on phase of the mass-transfer
before reaching the period minimum
\citep{del07amcvnevolution,sol10amcvnreview}.

\section*{Acknowledgements}

This work was supported by JSPS KAKENHI Grant Number 21K03616.
The author is grateful to the ZTF team
for making their data available to the public.
We are grateful to Naoto Kojiguchi for
helping downloading the ZTF data.
This research has made use of the AAVSO Variable Star Index
and NASA's Astrophysics Data System.

Based on observations obtained with the Samuel Oschin 48-inch
Telescope at the Palomar Observatory as part of
the Zwicky Transient Facility project. ZTF is supported by
the National Science Foundation under Grant No. AST-1440341
and a collaboration including Caltech, IPAC, 
the Weizmann Institute for Science, the Oskar Klein Center
at Stockholm University, the University of Maryland,
the University of Washington, Deutsches Elektronen-Synchrotron
and Humboldt University, Los Alamos National Laboratories, 
the TANGO Consortium of Taiwan, the University of 
Wisconsin at Milwaukee, and Lawrence Berkeley National Laboratories.
Operations are conducted by COO, IPAC, and UW.

The ztfquery code was funded by the European Research Council
(ERC) under the European Union's Horizon 2020 research and 
innovation programme (grant agreement n$^{\circ}$759194
-- USNAC, PI: Rigault).

We acknowledge ESA Gaia, DPAC and the Photometric Science
Alerts Team (http://gsaweb.ast.cam.ac.uk/\hspace{0pt}alerts).

\section*{List of objects in this paper}
\xxinput{objlist.inc}

\section*{References}

We provide two forms of the references section (for ADS
and as published) so that the references can be easily
incorporated into ADS.

\renewcommand\refname{\textbf{References (for ADS)}}

\newcommand{\noop}[1]{}\newcommand{\hyphalt}{-}

\xxinput{ayjaph.bbl}

\renewcommand\refname{\textbf{References (as published)}}

\xxinput{ayj.bbl.vsolj}

\end{document}